\documentclass[aps,prd,twocolumn,showpacs,amsmath,amssymb]{revtex4-1}
\usepackage{amsmath}
\usepackage{graphicx}
\usepackage{subfigure}
\usepackage{epstopdf}
\usepackage{color}
\usepackage{adjustbox}
\usepackage{multirow}
\usepackage{setspace}
\usepackage{overpic}
\usepackage{tabularx}
\usepackage{amssymb}
\usepackage[driverfallback=dvipdfmx, bookmarksnumbered, pdfstartview=FitH,colorlinks,urlcolor=blue, citecolor=blue,linkcolor=blue] {hyperref}

\usepackage{bm}
\usepackage{rotating}
\usepackage{verbatim}
\usepackage[utf8]{inputenc}

\newcommand{\BESIIIorcid}[1]{\href{https://orcid.org/#1}{\hspace*{0.1em}\raisebox{-0.45ex}{\includegraphics[width=1em]{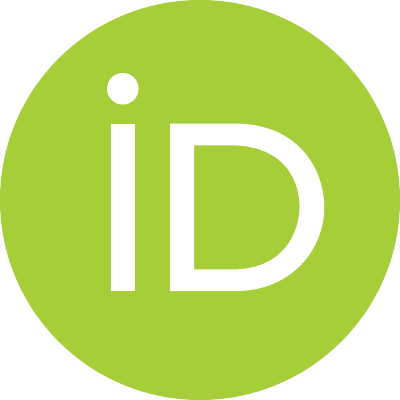}}}}

\begin{document}

\title{\bf \boldmath
  Search for the radiative decays $D^0\to \gamma \bar K_1(1270)^0$ and $D^+\to \gamma K_1(1270)^+$
}

\onecolumngrid

\author{
\centering
M.~Ablikim$^{1}$\BESIIIorcid{0000-0002-3935-619X},
M.~N.~Achasov$^{4,c}$\BESIIIorcid{0000-0002-9400-8622},
P.~Adlarson$^{81}$\BESIIIorcid{0000-0001-6280-3851},
X.~C.~Ai$^{86}$\BESIIIorcid{0000-0003-3856-2415},
R.~Aliberti$^{39}$\BESIIIorcid{0000-0003-3500-4012},
A.~Amoroso$^{80A,80C}$\BESIIIorcid{0000-0002-3095-8610},
Q.~An$^{77,64,\dagger}$,
Y.~Bai$^{62}$\BESIIIorcid{0000-0001-6593-5665},
O.~Bakina$^{40}$\BESIIIorcid{0009-0005-0719-7461},
Y.~Ban$^{50,h}$\BESIIIorcid{0000-0002-1912-0374}
% ...
\\
\vspace{0.2cm}
(BESIII Collaboration)\\
\vspace{0.2cm} {\it
$^{1}$ Institute of High Energy Physics, Beijing 100049, People's Republic of China\\
% ...
$^{4}$ Budker Institute of Nuclear Physics SB RAS (BINP), Novosibirsk 630090, Russia\\
% ...
$^{39}$ Johannes Gutenberg University of Mainz, Johann-Joachim-Becher-Weg 45, D-55099 Mainz, Germany\\
$^{40}$ Joint Institute for Nuclear Research, 141980 Dubna, Moscow region, Russia\\
% ...
$^{50}$ Peking University, Beijing 100871, People's Republic of China\\
% ...
$^{62}$ Southeast University, Nanjing 211100, People's Republic of China\\
% ...
$^{64}$ State Key Laboratory of Particle Detection and Electronics, Beijing 100049, Hefei 230026, People's Republic of China\\
% ...
$^{77}$ University of Science and Technology of China, Hefei 230026, People's Republic of China\\
% ...
$^{80}$ University of Turin and INFN, (A)University of Turin, I-10125, Turin, Italy; (B)University of Eastern Piedmont, I-15121, Alessandria, Italy; (C)INFN, I-10125, Turin, Italy\\
$^{81}$ Uppsala University, Box 516, SE-75120 Uppsala, Sweden\\
% ...
$^{86}$ Zhengzhou University, Zhengzhou 450001, People's Republic of China\\
\vspace{0.2cm}
$^{\dagger}$ Deceased\\
% ...
$^{c}$ Also at the Novosibirsk State University, Novosibirsk, 630090, Russia\\
% ...
$^{h}$ Also at State Key Laboratory of Nuclear Physics and Technology, Peking University, Beijing 100871, People's Republic of China\\
% ...
}
}

\begin{abstract}
 A search for the radiative decays $D^0\to \gamma \bar K_1(1270)^0$ and $D^+\to \gamma K_1(1270)^+$
is conducted using $20.3~\mathrm{fb}^{-1}$ of $e^+e^-$ annihilation data collected at the center-of-mass energy $\sqrt{s}=3.773$ GeV by the BESIII detector operating at the BEPCII collider.
No significant signals are observed, and upper limits on the branching fractions of
  $D^0\to \gamma \bar K_1(1270)^0$ and $D^+\to \gamma K_1(1270)^+$ at 90\% confidence level
  are  determined to be $7.7\times10^{-4}$ and $3.9\times10^{-5}$, respectively.
  This represents the first test of the Vector Meson Dominance mechanism in the radiative decays of charmed mesons to axial-vector mesons.
\end{abstract}

\maketitle

\oddsidemargin  -0.2cm
\evensidemargin -0.2cm

\section{Introduction}
\begin{figure}[htpb]
  \centering
  \includegraphics[width=0.224\textwidth]{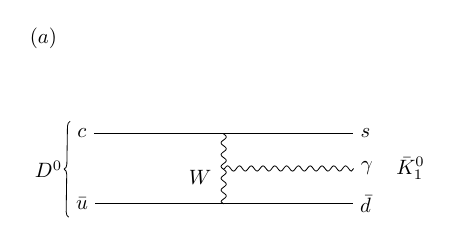}  
  \includegraphics[width=0.244\textwidth]{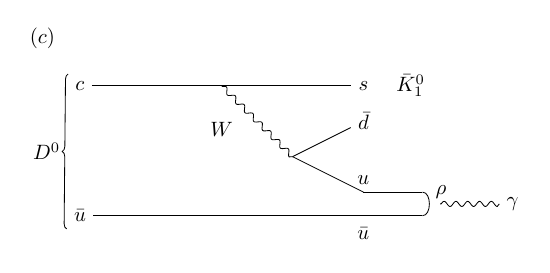}\\
   \includegraphics[width=0.224\textwidth]{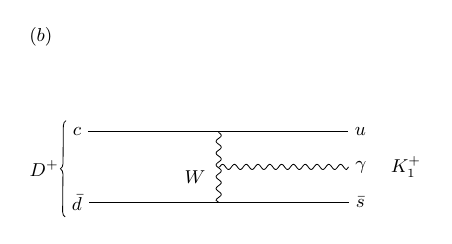}
   \includegraphics[width=0.244\textwidth]{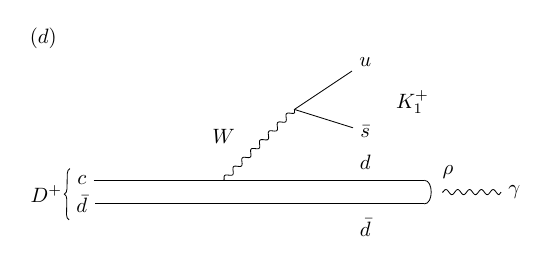}
  \caption{Feynman  diagrams for $D^{0} \to \gamma  \bar{K}_{1}^{0}$ and $D^{+} \to \gamma K_{1}^+$ decays, which are mediated via (a)(b)~short-distance processes and  (c)(d)~long-distance VMD processes.}
  \label{fig:feynman}
\end{figure}

In the Standard Model, short-distance contributions in  weak radiative decays of charmed hadrons are expected to be negligible compared to 
long-range processes. Long-range processes or potential contributions from new physics could enhance the branching fractions (BFs) of these decays by 2 to 3 orders of magnitude relative to short-range interactions~\cite{PhysRevD.52.6383,Fajfer:2015zea}.
Examples of short-distance and
 long-distance Vector Meson Dominance (VMD) processes for the decays $D^{0} \to \gamma \bar{K}_{1}^{0}$, $\bar{K}_{1}^{0} \to K^- \pi^+ \pi^0$ and $D^{+} \to \gamma K_{1}^+$, $K_{1}^{+} \to K^+ \pi^+ \pi^-$ are shown in Fig.~\ref{fig:feynman}. 
Throughout this paper, $\bar K^0_1$ and $\bar K^+_1$  denote the $\bar{K}_{1}(1270)^{0}$ and $K_{1}(1270)^+$ mesons, respectively.
The BFs of the decays
$D_{s}^+\to\gamma \rho^+$,
$D^{+} \to \gamma \rho^+$,
$D^{+} \to \gamma K^{*+}$,
$D^0\to\gamma \bar K^{*0}$,
$D^0\to\gamma\rho^0$,
$D^0\to\gamma\omega$,
and $D^0\to\gamma\phi$
have been measured by the BESIII~\cite{BESIII:gpt,BESIII:2024zfv},
Belle~\cite{Belle:2003vsx,Belle:2016mtj}, BaBar~\cite{BaBar:2008kjd}, CLEO~\cite{CLEO:1998mtp} collaborations.
However, the decays  $D^{0} \to \gamma  \bar{K}_{1}^{0}$ and $D^{+} \to \gamma K_{1}^+$  have not been observed. 
 The  BF  for the $D^+ \to \gamma K_{1}^+ $ decay mode is  predicted to be   $[(1.3 \pm 0.3), (1.5\pm 0.4)]\times10^{-5}$~\cite{2019Dlilun}. There  is  no specific BF predicted for  the $D^0$  channel, while its BF is constrained to be ${\cal O}(10^{-6})-{\cal O}(10^{-4})$~\cite{PhysRevD.52.6383}.

In this paper, we report the first search for the radiative decays $D^{0} \to \gamma  \bar{K}_{1}^{0}$ and $D^{+} \to \gamma K_{1}^+$ 
through the analysis of $e^+e^-$ annihilation data at the center-of-mass (c.m.) energy of 3.773 GeV with the BESIII detector corresponding to an integrated luminosity of 20.3 fb$^{-1}$~\cite{BESIII:2024lbn}. Charge conjugation is implied throughout this paper.

\section{Data and Monte Carlo}

The BESIII detector~\cite{BESIII:2009fln} records the symmetric $e^+e^-$ collisions
provided by the BEPCII storage ring~\cite{Yu:2016cof}
in the center-of-mass (c.m.) energy range from 1.84 to 4.95~GeV,
with a peak luminosity of $1.1 \times 10^{33}\;\text{cm}^{-2}\text{s}^{-1}$
achieved at $\sqrt{s} = 3.773\;\text{GeV}$.
BESIII has collected data samples corresponding to an integrated luminosity of 20.3 fb$^{-1}$ at the $\psi(3770)$ resonance~\cite{BESIII:2020nme,lu2020online,zhang:2022bdc}. The cylindrical core of the BESIII detector covers 93\% of the full solid angle and consists of a helium-based
multilayer drift chamber~(MDC), a plastic scintillator time-of-flight
system~(TOF), and a CsI~(Tl) electromagnetic calorimeter~(EMC),
which are all enclosed in a superconducting solenoidal magnet
providing a 1.0~T magnetic field. The solenoid is supported by an
octagonal flux-return yoke with resistive plate counter muon identification modules interleaved with steel. The charged-particle momentum resolution at $1~{\rm GeV}/c$ is
$0.5\%$, and the specific energy loss 
(${\rm d}E/{\rm d}x$)
resolution is $6\%$ for electrons
from Bhabha scattering. The EMC measures photon energies with a
resolution of $2.5\%$ ($5\%$) at $1$~GeV in the barrel (end cap)
region.
The time resolution in the TOF barrel region is 68~ps, while
that in the end cap region is 110~ps in 2015.
Then, the end cap TOF system was upgraded using multigap resistive plate chamber technology, providing a time resolution of 60~ps, which benefits 85\% of the data used in this analysis
~\cite{li2017study,guo2017study,Cao:2020ibk}.

Monte Carlo (MC) simulated data samples have been produced with a {\sc geant4}-based~\cite{GEANT4:2002zbu} software package. It includes the geometric description of the BESIII detector and the
detector response, and is used to determine the detection efficiencies and  estimate  backgrounds.
The simulation models include the beam-energy spread and initial-state radiation in the $e^+e^-$
annihilations  with the generator {\sc kkmc}~\cite{Jadach:2000ir,Jadach:1999vf}.
The inclusive MC samples consist of the production of $D\bar{D}$ pairs,
the non-$D\bar{D}$ decays of the $\psi(3770)$, the initial-state radiation
production of the $J/\psi$ and $\psi(3686)$ states and the
continuum processes. All particle decays are modelled with {\sc evtgen}~\cite{Lange:2001uf,Ping:2008zz} using the BFs taken from the
Particle Data Group~(PDG)~\cite{ParticleDataGroup:2024cfk},  when available,
or otherwise estimated with  {\sc lundcharm}~\cite{Chen:2000tv}. Final state radiation
from charged final state particles is incorporated using  {\sc photos} ~\cite{Richter-Was:1992hxq}.

 For the signal MC samples, we generated three million events for each of the decay channels shown in  Table~\ref{decay}. The radiative decays $D\to\gamma K_1$ are simulated using the HELAMP algorithm of EVTGEN~\cite{Ryd:2005zz} to ensure the conservation of angular momentum. The $K_1$ decays are simulated with the VSS\_PWAVE model~\cite{Ryd:2005zz}. 
 
\begin{table}
  \renewcommand{\arraystretch}{1.2}
  \centering
  \caption { The signal decays and their subsequent decays.}
  \renewcommand\arraystretch{1.2}
  \label{decay}
 
     \begin{tabular}{lll}
    \hline \hline
      Decay             & subdecay    & subdecay\\
      \hline
     \multirow{3}{*}{ $D^{0} \to \gamma \bar{K}_{1}^{0}$}      &   $\bar{K}_{1}^{0} \to \rho^+ K^-$    &        $\rho^+ \to \pi^+ \pi^0$              \\
  &   $\bar{K}_{1}^{0} \to K^{*-} \pi^+$               &     $K^{*-} \to K^- \pi^0$             \\
   & $\bar{K}_{1}^{0} \to \bar{K}^{*0} \pi^0$    &     $\bar{K}^{*0} \to K^- \pi^+$                \\
 \hline 

\multirow{2}{*}{ $D^{+} \to \gamma K_{1}^{+}$ }      &     $K_{1}^{+} \to K^+ \rho^0$   &            $\rho^0 \to \pi^+ \pi^-$           \\
  &    $K_{1}^{+} \to K^{*0} \pi^+$              &        $K^{*0} \to K^+\pi^-$             \\

    \hline \hline
  \end{tabular}
 
\end{table}

\section{Method}
At $\sqrt s=3.773$~GeV,    $D^+D^-$  or $D^0\bar{D}^0$ pairs are produced without any accompanying hadrons,
thereby offering a clean environment to investigate hadronic decays of the $D$ meson with a tag technique~\cite{MARK-III:1985hbd,BESIII:2023htx, Ke:2023qzc}. There are two types of samples selected with the tag technique: single-tag~(ST) and
double-tag~(DT)~methods. First, the ST method is employed to select a data sample with one $\bar D$ meson reconstructed. For the $D^{-}(\bar{D}^0)$ meson, the hadronic final states reconstructed are $K^+\pi^-\pi^-$, $K^0_S\pi^-$, $K^+\pi^-\pi^-\pi^0$, and $K^+K^-\pi^-$~($K^+\pi^-$, $K^+\pi^-\pi^0$, and $K^+\pi^+\pi^-\pi^-$). Subsequently, the DT method is utilized to constrain the process on the opposite side to search for the signal decays, $D^{+} \to \gamma K_{1}^{+}$ and $D^{0} \to \gamma  \bar{K}_{1}^{0}$.

The BF of a signal decay is determined by
\begin{equation}
  \label{eq:br}
  {\mathcal B}_{{\rm sig}} = N_{\rm DT}/(N^{\rm tot}_{\rm ST}\cdot\epsilon_{{\rm sig}}\cdot{\mathcal B}_{{\rm sub}}),
\end{equation}
where
$N^{\rm tot}_{\rm ST}=\sum_i N_{{\rm ST}}^i$ and $N_{\rm DT}$
are the total yields of the ST and DT candidates in data, respectively.
Here, ${\mathcal B}_{{\rm sub}}$ is the BF of the subsequent decay of $\bar{K_{1}^0} \to K^- \pi^+ \pi^0$  with $\pi^0 \to \gamma \gamma$ or $K_{1}^{+} \to K^+ \pi^+ \pi^-$.
The efficiency, $\epsilon_{{\rm sig}}$, for detecting the signal decay
is averaged over the tag modes $i$,
\begin{equation}
  \label{br}
  {
    {\mathcal \epsilon}_{\rm sig} = \frac{\sum_i (N^i_{\rm ST}\cdot \epsilon^i_{\rm DT}/\epsilon^i_{\rm ST})}{N^{\rm tot}_{\rm ST}},}
\end{equation}
where $\epsilon^i_{\rm ST}$ is the efficiency of reconstructing the ST mode $i$ (referred to as the ST efficiency), $N_{{\rm ST}}^i$ is the ST yield, and  $\epsilon^i_{\rm DT}$ is the efficiency of finding the tag mode $i$ and the  signal decay simultaneously. 

\subsection{ Tag Selection}
All charged tracks detected in the MDC must satisfy $\vert\!\cos\theta\vert<0.93$, where $\theta$ is defined as the polar angle with respect to the $z$ axis, which is the symmetry axis of the MDC. For charged tracks not originating from $K_S^0$ decays, the distance of closest approach to the interaction point (IP) is required to be less than 1\,cm in the transverse plane ($|V_{xy}|$), and less than 10\,cm along the $z$ axis ($|V_{z}|$). Particle identification (PID) for charged tracks combines the ${\rm d}E/{\rm d}x$ measurement in the MDC with the time of flight measurement of the TOF detector to define the likelihood function $\mathcal{L}(h)~(h=K,\pi)$ for each hadron ($h$) hypothesis. The charged kaons and pions are identified by comparing the likelihoods for the kaon and pion hypotheses, $\mathcal{L}(K)>\mathcal{L}(\pi)$ and $\mathcal{L}(\pi)>\mathcal{L}(K)$, respectively.

Each $K_{S}^0$ candidate is reconstructed from two oppositely charged tracks satisfying $|V_{z}|<$ 20~cm.
The two charged tracks are assigned as $\pi^+\pi^-$ without imposing further PID criteria.
They are constrained to originate from a common vertex and are required to have an invariant mass $M_{\pi^+\pi^-}$ such that $|M_{\pi^{+}\pi^{-}} - m_{K_{S}^{0}}|<$ 12~MeV$/c^{2}$, where $m_{K_{S}^{0}}$ is the known  $K^0_S$ mass~\cite{ParticleDataGroup:2024cfk}.
The decay length of the $K^0_S$ candidate is required to be greater than twice the vertex resolution away from the interaction point.
The $\chi^2$ of both the primary vertex fit and second vertex fit are required to be less than 100.

Photon candidates are selected by using the information recorded by the EMC. The time information of the crystal with the largest energy deposit in a cluster is required to be within [0, 700]\,ns of the event start time.

To exclude showers that originate from
charged tracks,
the opening angle  subtended by the EMC shower and the position of the closest charged track at the EMC
must be greater than  10 degrees as measured from the IP. 

The $\pi^0$ candidates are reconstructed from photon pairs each with an invariant mass within $(0.115,\,0.150)$\,GeV$/c^{2}$, which corresponds to about three times the invariant mass resolution.
 We remove $\pi^0$ candidates where both photons are not from the barrel region of the EMC to improve the resolution. The selected $\pi^0$ candidates are constrained to the known $\pi^0$ mass~\cite{ParticleDataGroup:2024cfk} via a kinematic fit to improve their energy and momentum resolution.

To separate the $D^-(\bar{D}^0)$ mesons from combinatorial backgrounds, two useful variables are defined. These are the beam-constrained mass, $M_{\rm BC}^{\rm tag}$, and the energy difference, $\Delta E_{\rm tag}$, which are defined as
\begin{eqnarray}
\begin{aligned}
		\Delta{E}_{\rm tag} &= E_{\rm beam}-E_{D^{-}(\bar{D}^0)},\\
    M_{\rm BC}^{\rm tag}&=\sqrt{E^2_{\rm beam}/c^4-|\vec{p}_{D^{-}(\bar{D}^0)}|^2/c^2},
 \label{eq:mbc}
\end{aligned}
\end{eqnarray}
where $E_{\mathrm{beam}}$ is the beam energy, and $E_{D^-(\bar{D}^0)}$ and $\vec{p}_{D^-(\bar{D}^0)}$ are the total energy and momentum of the $D^-(\bar{D}^0)$ candidate in the $e^+e^-$ center-of-mass (c.m.) frame, respectively.
If there is more than one  candidate in a given ST mode, the candidate with the smallest value of $|\Delta E_{\rm tag}|$ will be kept for the subsequent analysis.
The $\Delta E_{\rm tag}$ requirements and ST efficiencies are listed in Table~\ref{123}.

The ST yields are extracted by performing a binned maximum likelihood fit to the corresponding $M^{\rm tag}_{\rm BC}$ distribution. 

\begin{figure}[htbp]\centering
  \includegraphics[width=1.0\linewidth]{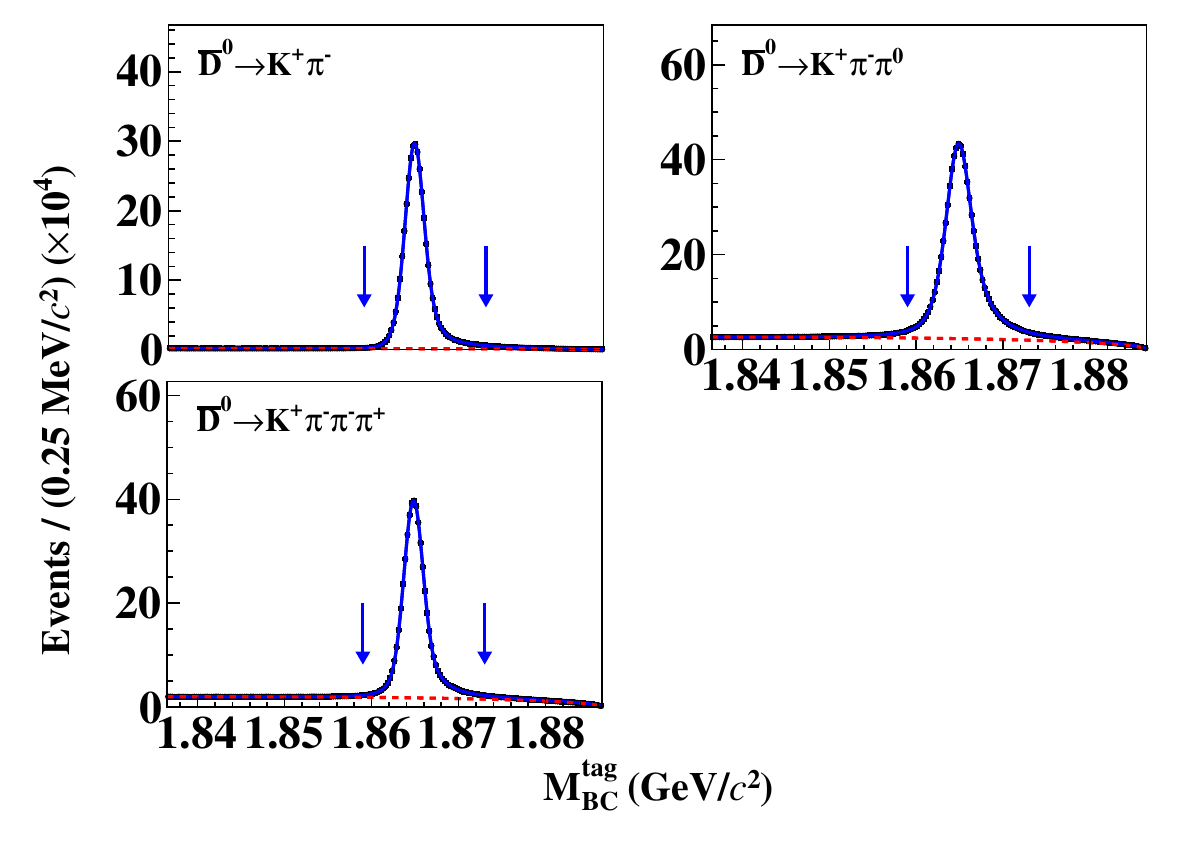}
  \includegraphics[width=1.0\linewidth]{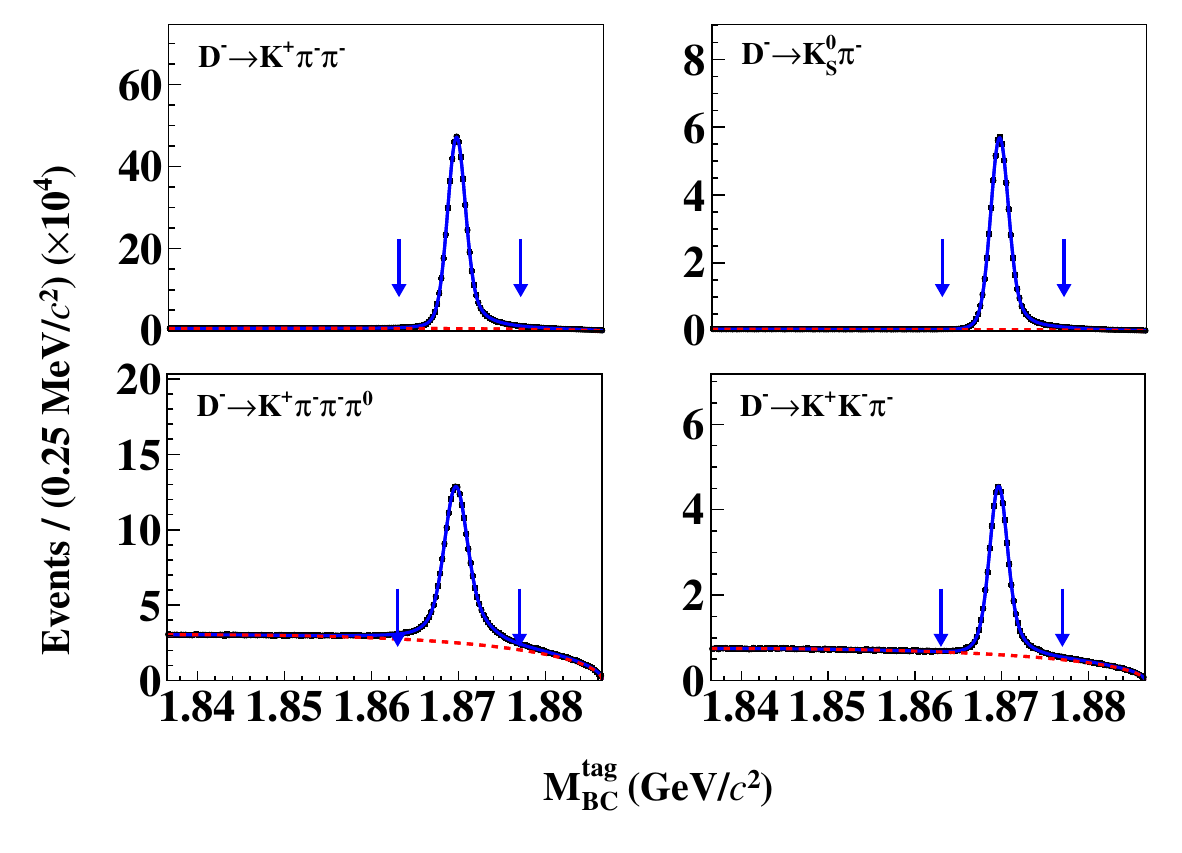}
  \caption{
    Fits to the $M^{\rm tag}_{\rm BC}$ distributions of the ST candidates for different tag modes.
    In each plot, the points with error bars are data, the blue curves are the best fits, and the red dashed curves describe the fitted combinatorial background shapes. The pair of blue arrows  indicate the signal window.}
    \label{figure:datafit_Massbc}
\end{figure}

 The signal shape is modeled by the sum of  two components. The first component is the shape determined from simulation convolved with a double-Gaussian function. The second component is a double-Gaussian.
  The background shape is described by the ARGUS function~\cite{ARGUS:1990hfq}, with the end-point parameter fixed at $E_{\rm beam}=1.8865$~GeV.~
Figure~\ref{figure:datafit_Massbc} shows the results of the  fits to the $M^{\rm tag}_{\rm BC}$ distributions of the accepted ST candidates  for different tag modes in data. The candidates with $M^{\rm tag}_{\rm BC}$ within~$(1.859,1.873)$~GeV/$c^2$ and $(1.863,1.877)$~GeV/$c^2$ are kept for the double-tag analysis for the $\bar{D^0}$ and $D^-$ channels, respectively.

\begin{table*}
\setlength{\tabcolsep}{8pt}
  \renewcommand{\arraystretch}{1.2}
  \centering
  \caption {The $\Delta E_{\rm tag}$ requirements, the $M^{\rm tag}_{\rm BC}$ requirements, the ST D meson yields in data~($N^i_{\rm ST}$), the ST efficiencies ($\epsilon_{\rm ST}^{i}$), and the DT efficiencies ($\epsilon_{\rm DT}^{i}$) for each tag mode. The uncertainties are statistical only.}
  \renewcommand\arraystretch{1.2}
  \label{123}
 
     \begin{tabular}{lccc@{ $\pm$ }ccc@{ $\pm$ }c}
    \hline \hline
    Tag mode                                      & $\Delta E_{\rm tag}$ (MeV) & $M^{\rm tag}_{\rm BC}$ (GeV/$c^2$)         & \multicolumn{2}{c}{$N^{i}_{\rm tag}$~$(\times 10^3)$}   &$\epsilon_{\rm ST}^i$ (\%) &\multicolumn{2}{c}{$\epsilon_{\rm DT}^i$}\\
    \hline
  
     $\bar{D^{0}} \to K^{+} \pi^{-}$                 & $(-0.27,0.27)$    &                \multirow{3}{*}{(1.859,\,1.873)}           & 3725.7&2.0   & $65.10\pm0.01$       &6.10&0.02   \\
      $\bar{D^{0}} \to K^{+} \pi^{-} \pi^{0}$                 & $(-0.62,0.49)$    &                  & 7422.3&3.2   & $35.60\pm0.01$   &2.81&0.01        \\
       $\bar{D^{0}} \to K^{+} \pi^{+} \pi^- \pi^-$                 & $(-0.26,0.24)$    &                  & 4987.5&2.5   & $40.94\pm0.01$     &3.18&0.01     \\
     \hline
    $D^{-} \to K^{+} \pi^{-} \pi^{-}$             & $(-0.25,0.24)$    & \multirow{4}{*}{(1.863,\,1.877)}               &  5552.8&2.5 & $51.10 \pm 0.01$  &8.46&0.01       \\
    $D^{-} \to K_{S}^{0} \pi^{-}$                 & $(-0.25,0.26)$    &                  & 656.5&0.8   & $51.42 \pm 0.01$        &8.71&0.04 \\
    $D^{-} \to K^{+} \pi^{-} \pi^{-} \pi^{0}$     & $(-0.57,0.46)$    &                  & 1723.7&1.8  & $24.40 \pm 0.01$       &3.28&0.01  \\
    $D^{-} \to K^{+} K^{-} \pi^{-}$               & $(-0.24,0.23)$    &      
    & 481.4&0.9   & $40.91 \pm 0.01$      &6.14&0.04    \\
   
    \hline \hline
  \end{tabular}
 
\end{table*}

\subsection{Signal Selection}

In the ST samples, candidates for  $D^{0} \to \gamma  \bar{K}_{1}^{0}$ and $D^{+} \to \gamma K_{1}^+$  are selected from the remaining tracks and showers not used in the tag-side reconstruction.
The $\bar{K_{1}^0}$ and $K_{1}^{+}$ mesons are reconstructed via $\bar{K_{1}^0} \to  K^- \pi^+ \pi^0$ and $K_1^{+} \to K^+ \pi^+ \pi^-$, respectively.
The $\pi^{\pm}$, $K^{\pm}$, $\pi^0$ and $\gamma$ candidates are selected with the same criteria as those used in ST selection.
The photon with the highest energy in the signal side is selected as the radiative photon.
To suppress combinatorial backgrounds and the backgrounds with $\pi^0$ meson, we require that there is no additional combination of two photons~($N^{\pi^0}_{\rm extra}$) that satisfy the $\pi^0$ selection criteria and no extra charged track~($N^{\rm charge}_{\rm extra}$) present in the event.
The invariant mass requirements of the $K_1$ candidates, $M_{K^- \pi^+ \pi^0}$ and $M_{K^+ \pi^-\pi^+}$, are optimized  with the Punzi method~\cite{Punzi:2020fsv} with the figure-of-merit (FoM) $\epsilon/(1.5+\sqrt{B})$. Here, $\epsilon$ is the signal efficiency based on the signal MC sample and $B$ is the background yield estimated from the inclusive MC sample.
Based on optimizations, we require the signal candidates to satisfy $M_{K^- \pi^+ \pi^0}$ within $(1.20,1.37)$ GeV/$c^2$ and $M_{K^+ \pi^-\pi^+}$ within $ (1.22,1.37)$ $\mathrm{GeV}/c^2$ for $D^0\to\gamma \bar{K}_{1}^{0}$ and $D^+\to\gamma K_{1}^{+}$, respectively.
Background from $D^0 (D^+) \to K_1^0(K_1^+) \pi^0$ decays, is suppressed through the use of the variable $M_{\gamma \gamma^{\rm extra}}$ which is the invariant mass of the radiative $\gamma$ and an extra $\gamma$  not used in the reconstruction of the tag and signal sides. Events that satisfy $M_{\gamma \gamma_{\rm extra}} < 0.115~{\rm GeV}/c^2$ or $M_{\gamma \gamma_{\rm extra}} > 0.15~{\rm GeV}/c^2$  are kept for further analysis.
To remove the background contamination from the decay $D^0 \to K^- \omega \pi^+$ in $D^0\to\gamma K_{1}^{0}$ decays and $D^+  \to K_{S}^0 K^+ \pi^0$ reconstructed as the $D^+ \to \gamma K_{1}^+ $ decay, we veto events with a recoil mass of $K^-\pi^+$ around the known $\omega$  mass~($0.75<M_{K^- \pi^+}^{\rm rec}<0.815$~GeV/$c^2$) and events with $\pi^+\pi^-$ invariant mass around the known  $K_S^0$ mass~($0.4677<M_{\pi^+ \pi^-}<0.5277$~GeV/$c^2$), respectively. 
For the signal side, we also  define the $\Delta{E}_{\rm sig}$ and $M_{\rm BC}^{\rm sig}$ in an analogous way to the tag side as in Eq.~\ref{eq:mbc}.

After optimization, we require the candidate events to satisfy $M_{\rm BC}^{\rm sig}$ within $ (1.862,1.869)~ \mathrm{GeV}/c^2$ for the $D^0\to\gamma \bar{K}_{1}^{0}$ channel and  $M_{\rm BC}^{\rm sig}$ within $ (1.867,1.874)~ \mathrm{GeV}/c^2$ for the $D^+\to\gamma K_{1}^{+}$ channel.

\section{Branching Fractions}

\begin{figure}[htpb]
  \centering  \includegraphics[width=1.0\linewidth]{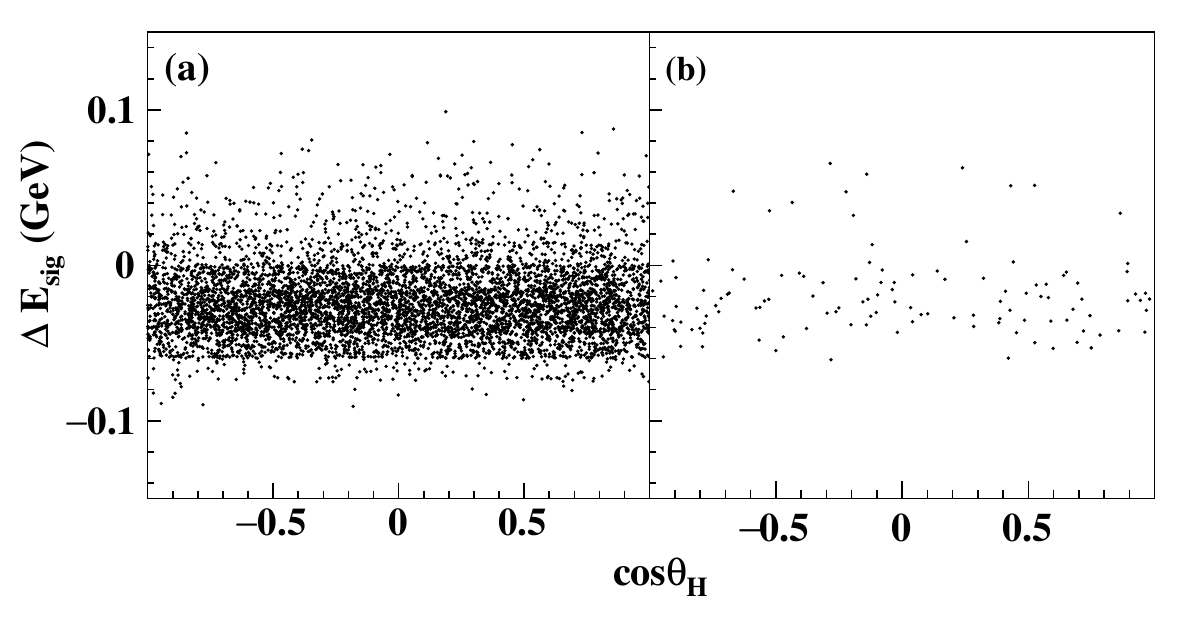}

  \caption{
    Distributions of $\Delta E_{\rm sig}$ versus $\cos \theta_{\rm H}$ in data of the DT candidate events for (a) $D^{0} \to \gamma  \bar{K}_{1}^{0}$~ and (b)$D^{+} \to \gamma K_{1}^+$~ decays.
    \label{plot:helix2d}
  }
\end{figure}

To further study combinatorial backgrounds and determine the signal yields, we define the $K_1$ helicity angle $\theta_H$~\cite{Berman:1965gi}, 
 which is the angle between the normal to the  $K_{1}$ decay plane, in the $K\pi\pi$ final state rest frame and the direction of flight of the $K_{1}$ meson in the $D$ meson rest frame.
The signal yields are extracted from a two-dimensional (2D) unbinned-maximum-likelihood fit to the $\Delta E_{\rm sig}$ versus $\cos \theta_{\rm H}$ distribution for the $D^{0} \to \gamma \bar{K}_{1}^{0}$ and $D^{+} \to \gamma K_{1}^+$ decays.
The $\Delta E_{\rm sig}$ ensures the property of a $D$ meson and $\cos\theta_{H}$ reveals that of $D \to \gamma K_1$.
The 2D distributions of  $\Delta E_{\rm sig}$ versus $\cos \theta_{\rm H}$ in data are shown in Fig.~\ref{plot:helix2d}. 
The dominant backgrounds for  the $D^{0} \to \gamma \bar{K}_{1}^{0}$ and  $D^{+} \to \gamma K_{1}^+$ decays are listed in Table~\ref{tab:bkg}.

\begin{table} 

  \centering
  \caption{
  The  dominant backgrounds  for the $D^{0} \to \gamma \bar{K}_{1}^{0}$ and $D^{+} \to \gamma K_{1}^+$ decay channels.  }
  \label{tab:bkg}
  \centering

    \begin{tabular}{lc}
      \hline \hline
            Dominant backgrounds      &   \% \\
      \hline
          $D^{0} \to K^- \pi^+ \pi^0 \pi^0$    & 84 \\
          $D^0 \to K^- \pi^0 \eta$ & 3 \\
          $D^0 \to K^- \pi^0 \omega$ &3\\
     
    \hline
    $D^{+} \to K^+ \pi^- \pi^+ \pi^0$ & 65\\
    $e^+e^-\to q\bar q$& 5\\
     $D^+ \to  K_S^0 \pi^+$&5\\
     $D^+ \to K^+ K_S^0 \pi^0$&3\\

       \hline \hline
   
    \end{tabular}
\end{table}

\begin{figure}[htpb]
  \centering
  \includegraphics[width=1.0\linewidth]{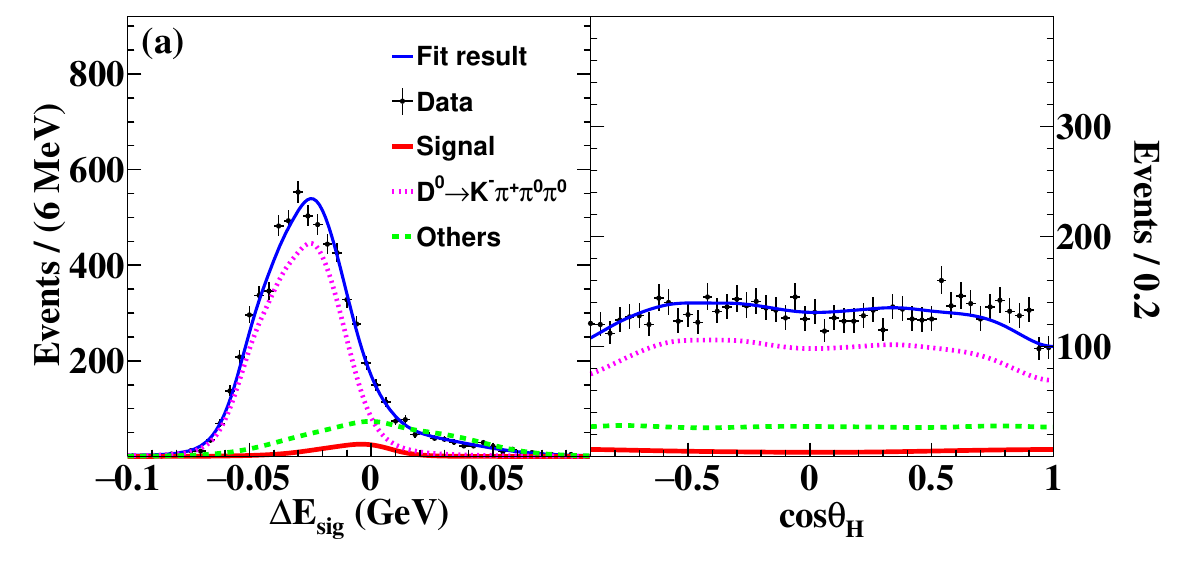} \\
 \includegraphics[width=1.0\linewidth]{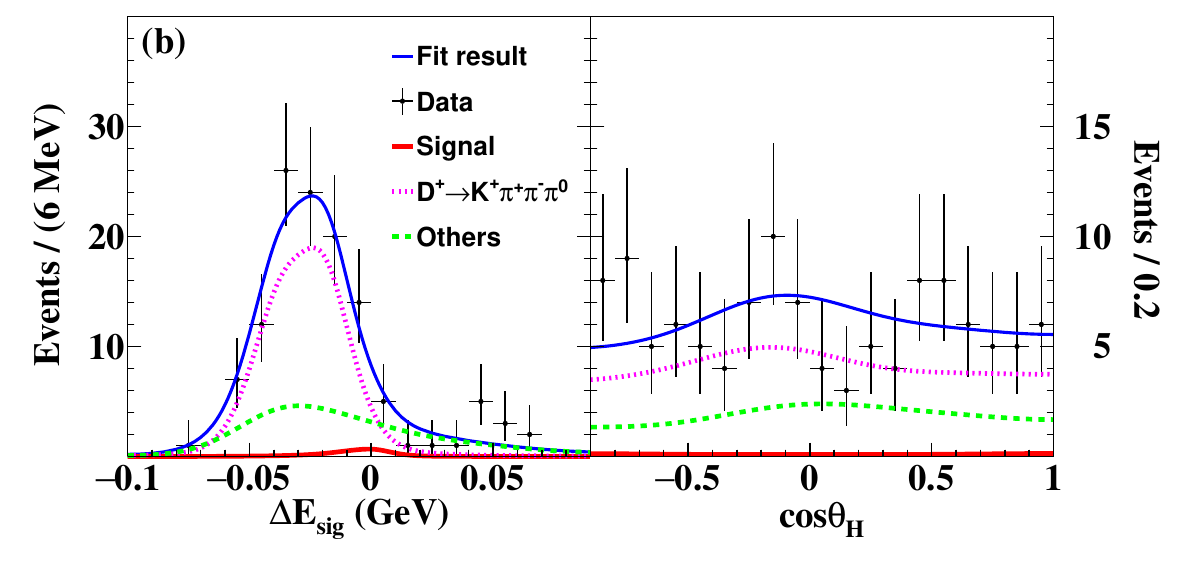}
  \caption{ One-dimensional projections of the 2D fits to the $\Delta E$ versus  $\cos \theta_{\rm H}$ distributions for the DT candidate events for
      $D^0\to\gamma \bar{K_{1}^0}$~(a) and $D^+ \to \gamma K_{1}^{+}$~(b).
    The dots with error bars are data, the blue solid curves are the fit results, the red solid curves are the  signal shapes, the pink dashed curves are the dominant background shapes and  the green dashed curves are  other background shapes. 
}
  \label{2Dfit1}
\end{figure}

\begin{figure}[htpb]
  \centering
  
  \includegraphics[width=0.5\textwidth]{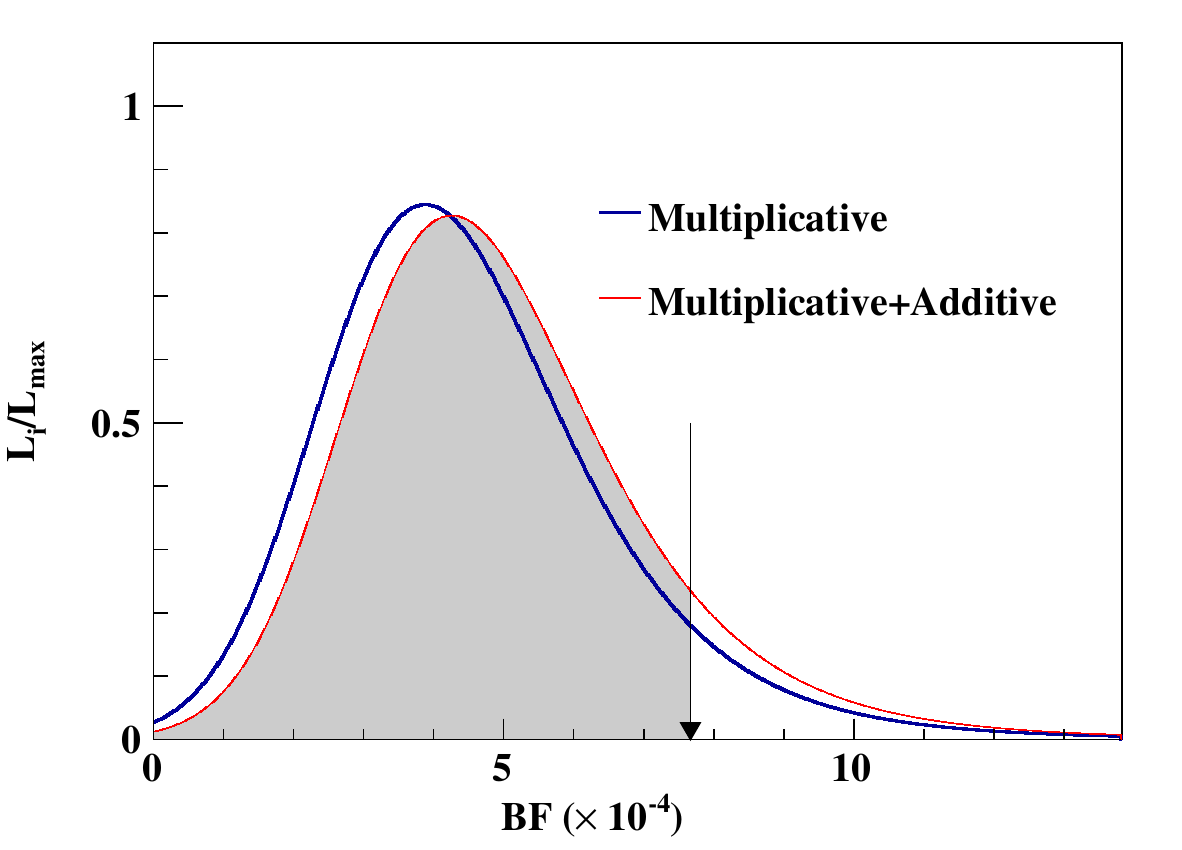}
  \includegraphics[width=0.5\textwidth]{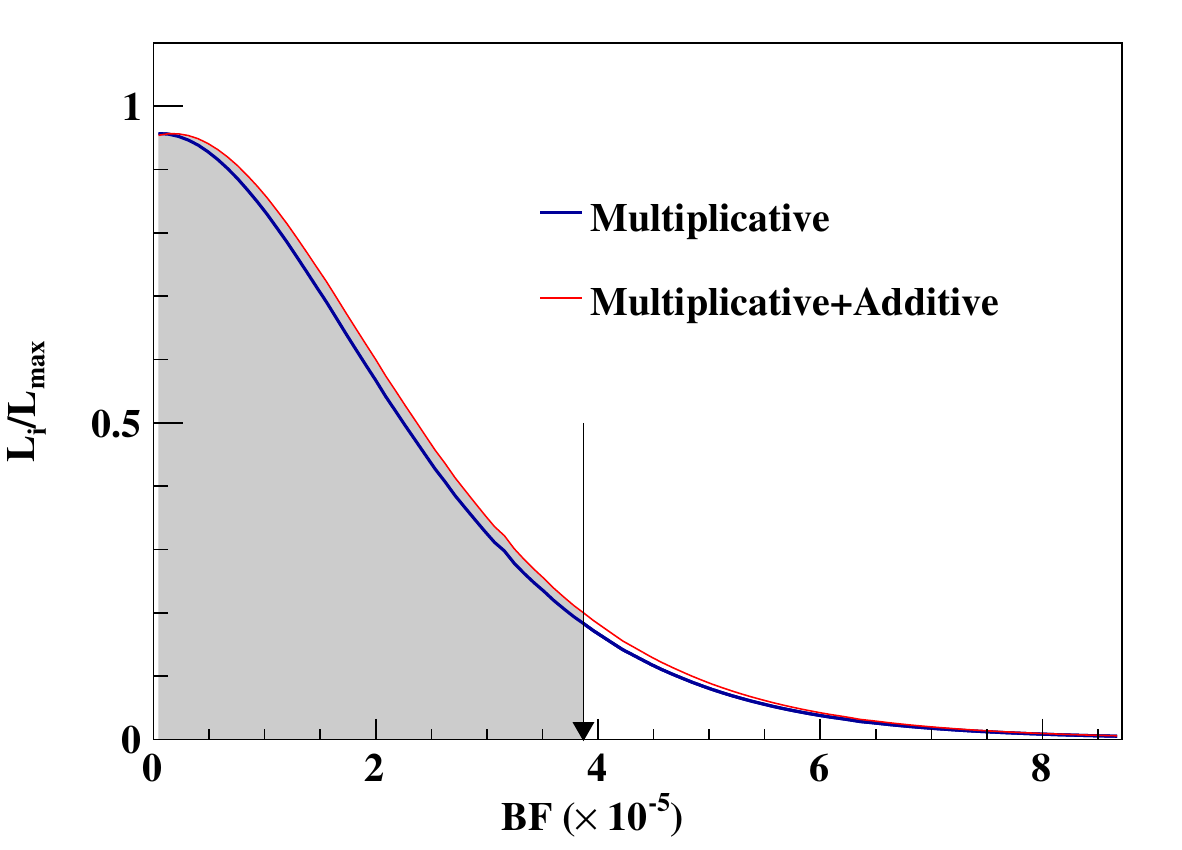}
  \caption{Scans of the likelihood functions versus the absolute BFs of $D^0\to\gamma \bar{K_{1}^0}$(up) and $D^+\to\gamma K_{1}^{+}$(down), obtained with the nominal fit model.
   The blue solid
 lines show the likelihood curves convolved with the multiplicative systematic uncertainty, while the red solid lines represent the likelihood
 curves that give the upper limit after incorporating the additive systematic uncertainty. The black arrows show the upper limits at 90\% confidence level.}
  \label{fig:upper}
\end{figure}

In the fits, the signal shapes are derived from the signal MC samples,
and the background shapes are derived from the inclusive MC samples with a Kernel 
Density Estimations technique~\cite{CRANMER2001198}.
Figure~\ref{2Dfit1} shows the fit projections for the $D^{0} \to \gamma \bar{K}_{1}^{0}$ and $D^{+} \to \gamma K_{1}^+$ decay channels.
The signal yields are
$199^{+69}_{-68}$ for the $D^{0} \to \gamma  \bar{K}_{1}^{0}$ decay
and
$2^{+8}_{-7}$ for the $D^{+} \to \gamma K_{1}^+$ decay.
Since no significant signal is found in either channel,
we set upper limits on the BF as shown in Fig.~\ref{fig:upper} by using the Frequentist approach~\cite{Stenson:2006gwf,Liu_2015} after considering the systematic uncertainties discussed later.
Finally, upper limits on the BFs of $D^{0} \to \gamma  \bar{K}_{1}^{0}$ and $D^{+} \to \gamma K_{1}^+$ are set to be $7.7\times10^{-4}$ and $3.9\times10^{-5}$ at the 90\% confidence level, respectively.

\section{Systematic uncertainty}
The sources of systematic uncertainties on the BF measurement are classified into two types: additive (or independent of the measured BF central value) and multiplicative (proportional to the BF).

 Additive uncertainties affect the signal yield determination, which is dominated by the imperfect background shape description. This systematic uncertainty is studied by altering the nominal MC background shape 
 with two methods.
  First, alternative MC samples are  used to derive  the background shape. Since the $D^0\to K^- \pi^+\pi^0\pi^0$ and $D^+\to K^+ \pi^+\pi^-\pi^0$ decays are the major background sources for $D^0\to\gamma \bar{K}_{1}^{0}$ and $D^+\to\gamma K_{1}^{+}$, we alter the MC shapes by varying the
 relative fractions of the background according to the PDG uncertainties. The largest change is taken as the corresponding systematic uncertainty. 
 Second, to account for the additive systematic uncertainty used to determine the background shape systematic uncertainties related to the fits,
the smoothing parameter in RooKeysPDF from RooFit~\cite{Cranmer:2000du, Verkerke:2003ir} used to implement the KDE is varied between 0 and 2.
Among the results of these fits, the largest upper limit on the BF is chosen. 

The multiplicative systematic uncertainties considered are described in the following. The uncertainty associated with the ST yield $N_{\rm ST}^{\rm tot}$, is assigned as 0.3\% after varying the signal and background shapes and floating the parameters of one Gaussian function in the fits~\cite{BESIII:2024slx}. Through studies of other channels ~\cite{BESIII:2024zfv} the systematic uncertainty in the efficiency due to tracking is determined to be 1.0\% for each charged track and the uncertainty due to the PID efficiency is 1.0\% for each $K^\pm$ or $\pi^\pm$ meson in the final state.
The systematic uncertainty due to the selection criteria of the $\gamma$ is studied with the $J/\psi\to \pi^+\pi^-\pi^0$ decay~\cite{BESIII:2011ysp}. The systematic uncertainty due to $\pi^0$ selection is examined through the DT hadronic $D\bar D$ events, as documented in Ref.~\cite{BESIII:2018mwk}.
These uncertainties are assigned as 1.0\% per photon and 2.0\% per $\pi^0$, respectively.
The differences in the $M^{\rm sig}_{\rm BC}$ resolution between data and MC simulation are obtained from the DT samples of  $D^0\to K^- \pi^+\pi^0\pi^0$ and $D^+ \to K^+ \pi^+\pi^-\pi^0$ decays reconstructed against the same tag modes used in the baseline analysis. The discrepancies in  efficiencies between data and MC simulation, 0.1\% and 1.2\%,  are taken as the systematic uncertainties.
The combined systematic uncertainty from the $N^{\pi^0}_{\rm extra}$ and $ N^{\rm charge}_{\rm extra}$ requirements are estimated to be 1.3\% and 0.4\%, which is also assigned by analyzing the control sample of $D^0\to K^- \pi^+\pi^0\pi^0$ and $D^+ \to K^+ \pi^+\pi^-\pi^0$ decays, respectively.
The used BFs of the $K_{1}^0 \to K^-\pi^+ \pi^0$  and $K_{1}^{+} \to K^+ \pi^-\pi^+$ decays are  $(46.2\pm9.3)\%$ and $(33.6\pm6.5)\%$ are taken from the world average~\cite{ParticleDataGroup:2024cfk}.
The systematic uncertainties associated with the MC model are estimated by changing the BFs of the known decay modes as well as the masses and widths 
 of the $\bar{K_1^{0}}$ and $K_1^{+}$ mesons.
Their uncertainties are dominated by the former source, and are 1.0\% and 0.6\%, respectively.

\begin{table}[h] 
  \centering
  \caption{
    Relative multiplicative systematic uncertainties (\%)   in the  BF  measurements.}
  \label{tab:relsysuncertainties}
  \centering

    \begin{tabular}{lcc}
      \hline \hline
     Multiplicative uncertainty                                       & $D^0 \to \gamma \bar{K}_{1}^{0}$ & $D^+ \to \gamma K_{1}^{+}$ \\
      \hline
      $N_{\rm ST}^{\rm tot}$                                     & 0.3                     & 0.3                     \\
      Tracking                                          & 2.0                     & 3.0                     \\
      PID                                               & 2.0                     & 3.0                     \\
      $\gamma$ and $\pi^0$ selection                    & 3.0                     & 1.0                     \\
      $M_{K \pi \pi}$ requirement                           & 0.2                     & 0.2                     \\
      $M_{\rm BC}$ requirement                          & 0.1                  &1.2                    \\
      $N_{\rm extra}^{\rm charge \& \pi^0} $ requirements & 1.3                    &0.4                     \\
      Quoted BFs                       &20.2                      &19.3 
                                 \\
    MC model                            &1.0 &0.6
    \\
      
      \hline
      Total                                             & 20.7                    & 19.8                    \\
       \hline \hline
   
    \end{tabular}
\end{table}

The total multiplicative systematic uncertainty is obtained by adding the individual components in quadrature.
Table~\ref{tab:relsysuncertainties} summarizes the sources of the systematic uncertainties in the BF measurements.

\section{Summary}

By analyzing a data sample corresponding to an integrated luminosity of 20.3~fb$^{-1}$ collected in $e^+e^-$ annihilations  at $\sqrt{s} = 3.773\;\text{GeV}$, the radiative decays $D^0\to \gamma \bar K_1(1270)^0$ and $D^+\to \gamma K_1(1270)^+$
are searched for for the first time. No significant signal is observed.
The upper limits on the BFs of the $D^{0} \to \gamma  \bar{K}_{1}^{0}$ and $D^{+} \to \gamma K_{1}^+$ decays at the 90\% confidence level are set to be
$7.7\times10^{-4}$ and $3.9\times10^{-5}$, respectively. These results are consistent with theoretical expectations.

\section{Acknowledgement}
The BESIII Collaboration thanks the staff of BEPCII (https://cstr.cn/31109.02.BEPC) and the IHEP computing center for their strong support. This work is supported in part by National Key R\&D Program of China under Contracts Nos. 2023YFA1606000, 2023YFA1606704; National Natural Science Foundation of China (NSFC) under Contracts Nos. 12405094, 11635010, 11935015, 11935016, 11935018, 12025502, 12035009, 12035013, 12061131003, 12192260, 12192261, 12192262, 12192263, 12192264, 12192265, 12221005, 12225509, 12235017, 12361141819; Joint Large-Scale Scientific Facility Fund of the NSFC and the Chinese Academy of Sciences under Contract No.~U2032104; the Excellent Youth Foundation of Henan Scientific Committee under Contract No.~242300421044; 
Science and Technology R$\&$D Program Joint Fund Project of Henan Province  (Grant No.225200810030); Science and Technology Innovation Leading Talent Support Program of Henan Province; 
Program of Science and Technology Development Plan of Jilin Province of China under Contract No. 20230101021JC;
The Chinese Academy of Sciences (CAS) Large-Scale Scientific Facility Program; the Strategic Priority Research Program of Chinese Academy of Sciences under Contract No. XDA0480600; CAS under Contract No. YSBR-101; 100 Talents Program of CAS; The Institute of Nuclear and Particle Physics (INPAC) and Shanghai Key Laboratory for Particle Physics and Cosmology; ERC under Contract No. 758462; German Research Foundation DFG under Contract No. FOR5327; Istituto Nazionale di Fisica Nucleare, Italy; Knut and Alice Wallenberg Foundation under Contracts Nos. 2021.0174, 2021.0299; Ministry of Development of Turkey under Contract No. DPT2006K-120470; National Research Foundation of Korea under Contract No. NRF-2022R1A2C1092335; National Science and Technology fund of Mongolia; Polish National Science Centre under Contract No. 2024/53/B/ST2/00975; STFC (United Kingdom); Swedish Research Council under Contract No. 2019.04595; U. S. Department of Energy under Contract No. DE-FG02-05ER41374.

\bibliography{refer}

\end{document}